# Attomicroscopy imaging and control of electron motion in graphene

*Mingrui Yuan [1,2][†], Husain Alqattan [1][†], Dandan Hui[1], Mohamed Sennary[1], Vladimir Pervak[3], Nikolay V. Golubev\* [1] and Mohammed Th. Hassan\* [1,2]*

[1] Department of Physics, University of Arizona, Tucson, AZ 85721, USA.

[2] James C. Wyant College of Optical Sciences, University of Arizona, Tucson, Arizona 85721, USA.

[3] Ludwig-Maximilians-Universität München, Am Coulombwall 1, 85748, Garching, Germany.

[†] These authors contributed equally to this work

\*Corresponding author e-mail: mohammedhassan@arizona.edu and ngolubev@arizona.edu

## Abstract

Attosecond science has leveraged the highly nonlinear interactions between intense few-cycle laser pulses and matter, allowing for unprecedented observation and control of electron motion with remarkable temporal resolution. However, most existing experiments focusing on laser-controlled attosecond dynamics have dealt with quasi-bound electrons released in the ionization continua of atoms, molecules, or conduction bands in solid-state systems. Here, we employed the recently developed attomicroscopy imaging tool to investigate, visualize, and manipulate the motion of bound electrons in graphene. By adjusting the carrier-envelope phase and the field strength of the driving electric field, we were able to control both the amplitude and direction of the field-induced electron current between carbon atoms in graphene. This research opens new avenues for understanding and controlling dynamic, on-demand electron motion processes, including chemical reactions, molecular bonding, and the electronic properties of materials.

## 1. Introduction

The control of electron motion in matter is a long-standing goal of attosecond physics. The required prerequisite to manifest such control is the necessity to probe or image the electron motion in real time and space. At the beginning of the millennium, the generation of attosecond extreme ultraviolet (EUV) pulses—based on the strong field interaction in a high harmonic generation (HHG) process—gave birth to the attosecond science[1]. Accordingly, the development of attosecond stroboscopic spectroscopy paved the way to probe the electron dynamics in ionized systems[2-10]. Hence, this electron tracing capability has been successfully used to control the free electron wave packets on the attosecond time scale by tuning the carrier-envelope phase (CEP) of the laser, as demonstrated earlier[11], where the signatures of the control were printed and retrieved from the measured EUV spectra. Other important demonstrations of electron motion in the continuum were reported in the HHG process by using a polarization-gated laser pulse[12] and two colour laser pulses[13]. Following this work, the steering of the electron wave packets by light field has been utilized to sample the light field waveform by attosecond streaking[14-17] and to determine the internal delay in the photoemission process[18]. In addition, the light-induced coherent electron motion and hole density distribution in ionized atoms have been controlled by synthesized light transients[19]. Later, similar free electron motion control has been demonstrated in solid-state systems[20-25] and molecules[26-28]. The generation of an attosecond optical laser pulse made it possible to probe and control the bound electrons in neutral atoms in the gas phase[29]. Furthermore, the light-induced current due to the collective electron motion dynamics has been controlled by changing the waveform and the CEP in the solid state[30-32]. Recently, we were able to control the electron motion in dielectric systems by synthesized attosecond light fields and by changing the intensity of the driver field[33-35].

Nonetheless, despite these impressive advancements in the field of attosecond physics, the bound electron motion control in nanostructures in spatiotemporal dimensions remains limited. The electron dynamics is strongly entangled with the system structure in the space domain. Hence, access and electron motion control in both real-time and space domains are crucial for realistic engineering applications such as the development of ultrafast optical electronics devices [36].

Here, we demonstrate the field-driven bound electron motion control in graphene nanostructure probed by Attomicroscopy diffraction imaging[37]. Although the interpretation of the time-resolved diffraction measurements represents a notable difficulty in comparison to those performed with a stationary target, the recently developed theory of attosecond diffraction imaging[38] makes it possible to retrieve the time-dependent electron current direction and amplitude between the carbon atoms in the real space of graphene. We, therefore, report a combined theoretical and experimental study of the ultrafast electron dynamics in graphene sample and demonstrate the possibility to control the electron motion in nanostructures by appropriately tailored laser pulses. Besides controlling the graphene electron motion dynamics in real-time, we demonstrate the control of the electron motion's direction by changing the CEP of the driver field and the electron momentum's amplitudes by changing the intensity of the field.

## 2. Laser field-driven electron dynamics in graphene

In the regime of strong field interaction of pump laser with graphene, the electrons can be excited from the valence band to the conduction band by photon absorption or through the Landau–Zener mechanism at the vicinity of Dirac points, as illustrated in Figure 1. The excited electrons in the conduction band and the corresponding created holes in the valence band continue experiencing the action of the driver field, thus accelerating and decelerating in their bands according to the Bloch acceleration theorem $\mathbf{k}_t = \mathbf{k} + e\mathbf{A}(t)/\hbar$. Here, $\mathbf{k}$ and $\mathbf{k}_t$ are field-free and field-accelerated momenta, respectively, of charge carriers placed in the external electric field $\mathbf{E}(t)$ that has the corresponding vector potential $\mathbf{A}(t) = -\int_{-\infty}^{t} \mathbf{E}(t')dt'$. Therefore, one should expect that the electron momentum distribution (EMD) will be displaced in the reciprocal space of graphene depending on the instantaneous value of the vector potential amplitude.

When the vector potential amplitude is negative and polarized along the X axis [Figure 1a(I)], the EMD is displaced towards the negative direction ($K_{-X}$, Figures 1b and 1c). Importantly, the change of the EMD in the reciprocal space will be reflected in the real space electron densities and, in addition, will generate the electric current in a positive +X direction (common convention is to define current as moving in the opposite direction as the negative charge flow), as illustrated in Figure 1d. As the field evolves in time by a quarter field cycle (typically in the subfemtosecond range for the near-infrared (NIR) driver laser pulse), the value of the vector potential becomes zero (Figure 1a(II)). In this case, the excited carriers in the reciprocal space of graphene remain unaffected by the field (Figure 1e). Accordingly, the EMD is nearly symmetric around the Dirac point, as shown in Figure 1f. Thus, there is no raised electron current between carbon atoms (Figure 1g). When the vector potential becomes positive (Figure 1a(III)), the excited carriers are accelerated to the other side in the reciprocal space by gaining positive momentum from the driver field (see Figure 1h), and EMD becomes displaced towards the positive direction ($K_{+X}$), as depicted in Figure 1i. In this case, the generated current flows towards -X direction, as illustrated in Figure 1j. Hence, the direction and amplitude of electron motion in graphene are alternated in the presence of the strong laser field depending on the instantaneous value of the field and its intensity. Accordingly, it is extremely appealing to explore the possibilities of controlling the electron motion dynamics between carbon atoms in graphene on-demand by varying the waveform and intensity of the driver laser field as we present next.

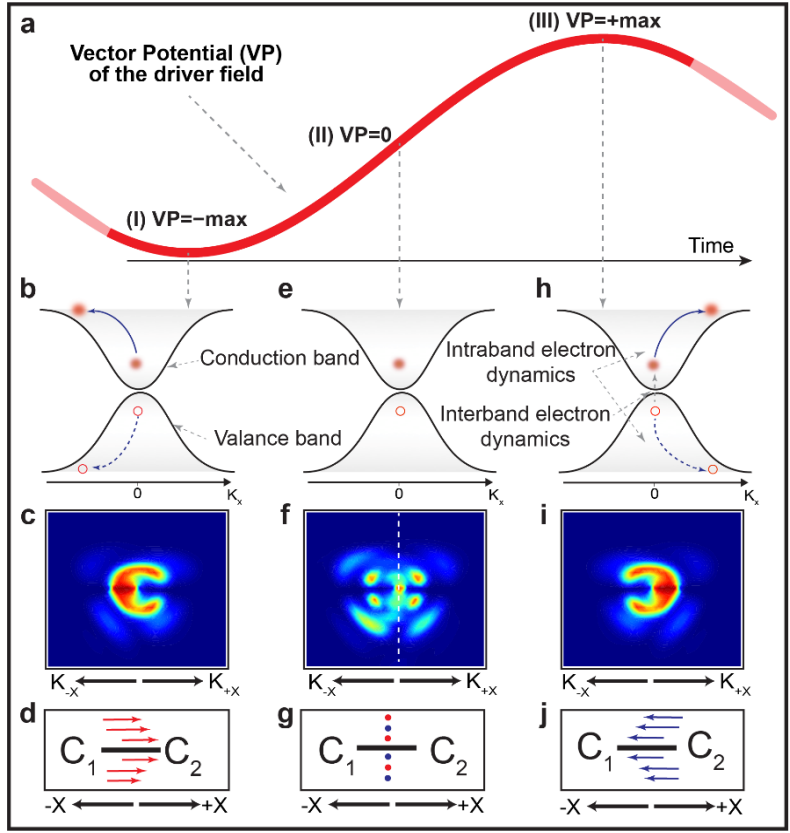

**Figure 1. The field-driven electron motion in graphene. (a) The driver field changes from -max, 0, to +max amplitude values as it evolves over time. (b-d), (e-g) and (h-j) show the electron-hole motion in the valence and conduction bands, the excited carriers' distribution in reciprocal space, and the electron current direction between graphene's carbon atoms at different instances/amplitude values of the driver field.**

### 3. Imaging the electron motion in graphene

The field-driven electron motion in solids can be probed by electron diffraction with a temporal resolution higher or equal to the sub-cycle field duration. It has been shown in our recent studies[37,38] that the intensities of scattered beams can be expressed as:

$$I(\mathbf{S},t) = \frac{1}{N} \sum_{n,m,f} \int_{\text{unit cell}} \rho_{n,m}(\mathbf{k},t) \hat{F}_S^*[Q_{f,m}(\mathbf{k}_t,\mathbf{r})] \hat{F}_S[Q_{f,n}(\mathbf{k}_t,\mathbf{r})] d\mathbf{k}, \qquad (1)$$

where **S** is the scattering vector, **r** is the real space coordinate of an electron, $\rho_{m,n}(\mathbf{k},t)$ are the elements of the electron density matrix in the reciprocal space, $Q_{i,j}(\mathbf{k},\mathbf{r})$ are the transition electron densities connecting the bands with each other, $\hat{F}_S$ is the Fourier transform operator, and N is the normalization constant. As one can see from Eq. (1), the scattering intensities depend on the real-space properties of the involved electronic bands via $Q_{i,j}(\mathbf{k},\mathbf{r})$ and on the EMD in these bands via $\rho_{m,n}(\mathbf{k},t)$. Accordingly, by measuring the diffraction intensities as a function of pump-probe delay, one can access the microscopic properties of a system such as the EMD and eventually connect it with observable quantities such as the electron density and the generated electric current. In the following, we use our recently developed attomicroscope setup[37] that provides sub-femtosecond imaging resolution to probe the bound electron motion dynamics in graphene by means of electron diffraction.

In the performed experiment, initially, the laser beam of a linearly polarized few-cycle pulse—carried at centre wavelength ~780 nm and with passively stabilized CEP—is divided into two beams. Then, the first beam passes through a neutral density filter to control its power, and it is utilized as a pump pulse. The second beam goes through a polarization-gating process to generate a polarization-gated half-cycle pulse[37]. This beam is sent to the attomicroscope to generate attosecond electron pulses by optical gating approach [37], which are used to probe the field-induced electron motion of multilayer (single crystal) graphene that is triggered by the pump pulse. The pump field strength is set to be ~2.5 V/nm, and it is linearly polarized (p-polarization) along the bond between the carbon atoms in the graphene sample. We measured the graphene's diffraction pattern (see inset in Figure 2a) as a function of the time delay between the pump pulse and the attosecond probe gated electron pulse. The measured change of the intensities of first-order Bragg peaks as a function of pump-probe delay is plotted in black dots in Figure 2a.

To understand the nature of the measured diffraction modulations and connect it with the properties of the system, we calculated the field-driven evolution of electron density of graphene in reciprocal space using the semiconductor Bloch equation:

$$i\hbar \frac{\partial}{\partial t} \rho_{n,m}(\mathbf{k},t) = \left[ E_m(\mathbf{k}_t) - E_n(\mathbf{k}_t) \right] \rho_{n,m}(\mathbf{k},t) \\ + \mathbf{E}(t) \cdot \{ \mathbf{D}(\mathbf{k}_t), \boldsymbol{\rho}(\mathbf{k},t) \}_{n,m} - i \frac{1-\delta_{n,m}}{T_d} \rho_{n,m}(\mathbf{k},t), \qquad (2)$$

where $E_m(\mathbf{k})$ and $E_n(\mathbf{k})$ are band energies, $\mathbf{D}(\mathbf{k}) = \left(\mathbf{D}_x(\mathbf{k}), \mathbf{D}_y(\mathbf{k}), \mathbf{D}_z(\mathbf{k})\right)^T$ is a vector of matrices of the transition dipole moments connecting the bands with each other, the commutator symbol "{}" is defined as $\{A, B\} = AB - BA$, and $T_d$ is the interband dephasing time. Solving Eq. (2) for a given electric field $\mathbf{E}(t)$ and with the given initial conditions (all the electrons stay in the valence band before the interaction with the field), one obtains the electron density matrix $\rho(\mathbf{k}, t)$ whose diagonal elements represent the EMD in the corresponding bands while off-diagonal elements describe electronic coherences.

Although some parameters of the driver field $\mathbf{E}(t)$, such as the photon energy and the intensity, are known from the experimental setup design, it is not always fully clear in advanced ultrafast experiments what waveform and CEP the applied electric field has when it reaches the sample. Therefore, we extracted the required field parameters from the measured diffraction signal by fitting a Gaussian-shaped single-frequency waveform to the experimental data (the fitting process is explained in Supporting Information). We assumed that the retrieved function (plotted by the red line in Figure 2a) is proportional to the vector potential $\mathbf{A}(t)$ of the applied field $\mathbf{E}(t)$. Accordingly, we used the obtained electric field to calculate the reciprocal space density matrix $\rho(\mathbf{k}, t)$ by numerically solving Eq. (2) in the graphene unit cell employing the two-level tight-binding model (see Ref. [16] for more details).

The obtained reciprocal space density matrix $\rho(\mathbf{k}, t)$ gives access to various properties of the system such as the aforementioned EMDs and, together with the electron densities $Q_{i,j}(\mathbf{k}, \mathbf{r})$ retrieved from the tight-binding model, allows one to calculate the diffraction intensities via Eq. (1). The computed diffraction intensity of first-order Bragg peaks of graphene is depicted by the blue solid line in Figure 2a demonstrating very good agreement with the measured signal. In addition, $\rho(\mathbf{k}, t)$ is an intermediate quantity that links the measured diffraction signal with the real-space properties of the system, such as the electron density:

$$Q(\mathbf{r}, t) = \frac{1}{N} \sum_{n,m} \int_{\text{unit cell}} \rho_{n,m}(\mathbf{k}, t) Q_{m,n}(\mathbf{k}_t, \mathbf{r}) d\mathbf{k}, \qquad (3)$$

and the electron current density (ECD):

$$\mathbf{J}(\mathbf{r}, t) = \frac{\hbar/m}{N} \sum_{n,m} \int_{\text{unit cell}} \text{Im}\{\rho_{n,m}(\mathbf{k}, t) \nabla_\mathbf{r} Q_{m,n}(\mathbf{k}_t, \mathbf{r})\} d\mathbf{k} - \frac{e}{m} \mathbf{A}(t) Q(\mathbf{r}, t), \qquad (4)$$

which, in turn, is composed of the paramagnetic (first term in Eq. (4)) and diamagnetic (second term in Eq. (4)) contributions, and the relativistic effects are neglected. We, therefore, can establish the connection between the measured diffraction signal and the real-space properties of the system.

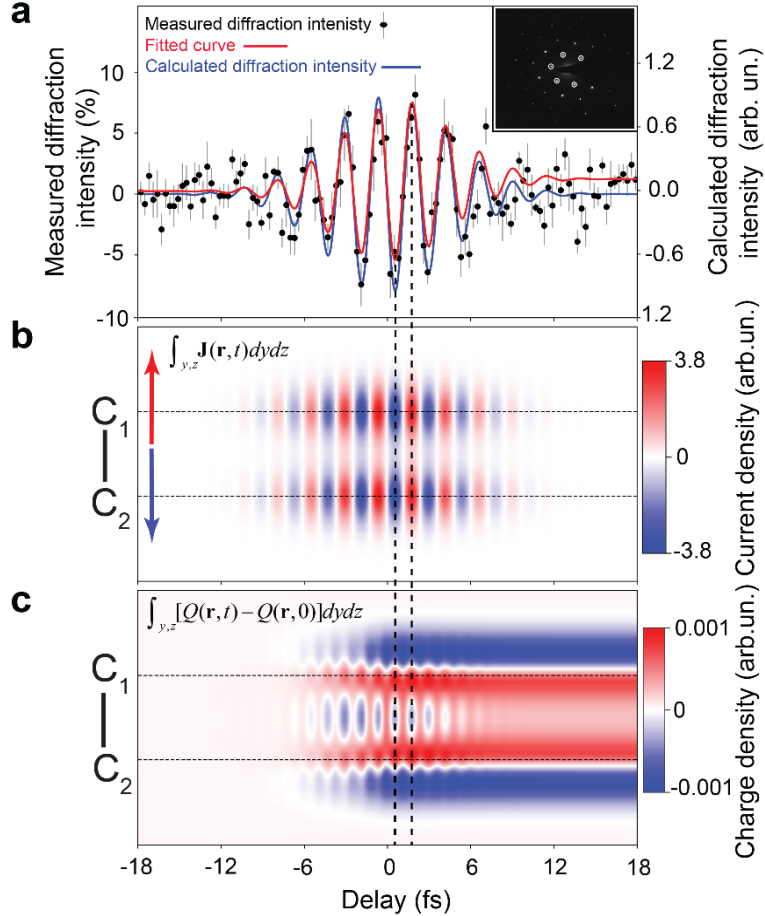

**Figure 2. Imagining the graphene's electron motion. (a) The measured (black dots; fitting is shown by a red solid line, and the error bars are retrieved from three diffraction dynamics scans) and calculated (blue solid line) diffraction intensities of the first-order Bragg peaks of graphene as a function of the delay between the driver field and the attosecond electron pulse. The inset shows the diffraction pattern of graphene. (b) The generated electron current between two carbon atoms ($C_1$ and $C_2$) of graphene. The red colour shows that the current is directed from $C_2$ to $C_1$, while blue represents the current direction from $C_1$ to $C_2$. The direction of the electron current is altered every half-cycle of the driver pulse. (c) The electron density between the two carbon atoms oscillates with double the frequency of the driver field.**

Figures 2b and 2c depict the real-space electron current ($\int_{y,z} \mathbf{J}(\mathbf{r},t) dy dz$) and electron difference ($\int_{y,z} [Q(\mathbf{r},t) - Q(\mathbf{r},0)] dy dz$) densities, respectively, integrated over coordinate axes perpendicular to the C-C bond of graphene. We note that the calculated diamagnetic current is found to be much stronger than the paramagnetic one, which was expected for a strong external field like the one used in our experiment. As we can see from Figure 2b, the ECD magnitude and direction alternate from $C_2$ to $C_1$ (presented by red colour code) to $C_1$ to $C_2$ (presented by blue colour code) every half-cycle of the driver field. When the vector potential is negative (at $\tau \approx 0$ fs), the electrons are pushed in the $C_2$ to $C_1$ direction, creating thus the electric current from $C_1$ to $C_2$ following the explanation presented in Section 2. After half field cycle (at $\tau \approx +1.2$ fs), the direction of the current flips to be from $C_2$ to $C_1$. Therefore, the oscillation period of the electric current matches that of the applied driver field. At the same time, the instantaneous electron densities at the minimum and maximum values of the vector potential are nearly identical to each other, as shown in Figure 2c. Accordingly, the oscillation period of the electron density dynamics is observed to be twice the frequency of the driver field. We note in passing that the scattering event implies the exchange of momenta between electrons of the beam and the target. Therefore, the same instantaneous electron density with opposite directions of the instantaneous momenta will scatter electrons differently (see Ref. [15] for a detailed discussion). This, in turn, is the main physics underlying the electron scattering from the time-dependent target in our experiment. Accordingly, we are able to extract the fingerprints of the electron motion from the measured diffraction pattern. In the next section, we will utilize these unique capabilities to monitor the direction and magnitude of electron currents and demonstrate the control over the microscopic electron motion in graphene with the attosecond time resolution.

**3. Controlling the electron motion in graphene**

Let us now explore the possibilities of controlling the electron motion in graphene by appropriately tailoring the driver laser field. We would like to demonstrate two scenarios for controlling the electron dynamics via (i) alternating the direction of electron motion by varying the CEP of the driver field and (ii) manipulating the amplitude of electron motion by changing the intensity of the field.

First, to demonstrate the ECD direction field control, we measured the scattering intensity of the first-order diffraction peaks of graphene as a function of time at CEP~0 and plotted it in Figure 3a (black dots, fitted curve is depicted by red solid line). Then, we implemented a 10 μm-thin SiO$_2$ dispersion element in the pump beam path to alter the CEP by $\pi$ and record the diffraction measurements, as shown in Figure 3b. Notably, the retrieved CEPs difference from the fitting in Figure 3a and Figure 3b is 1.07 $\pi$ (see Table 1 in Supporting Information), which agrees well with the estimated CEP change in our experiment. The corresponding calculated diffraction intensities for each field waveform are depicted by the blue solid lines in Figure 3a and b, which demonstrate a good agreement with the measured signals. Figure 3c and Figure 3d show the calculated ECD between the carbon atoms in graphene at $\tau = 0$ fs for the driver fields with CEP ~0 & $\pi$, respectively. Remarkably, the direction of electron motion for the two applied laser fields is completely reversed, which is reflected clearly in the measured scattering intensities. These results show the precise attosecond control of the electron current direction by altering the CEP of the driver laser field.

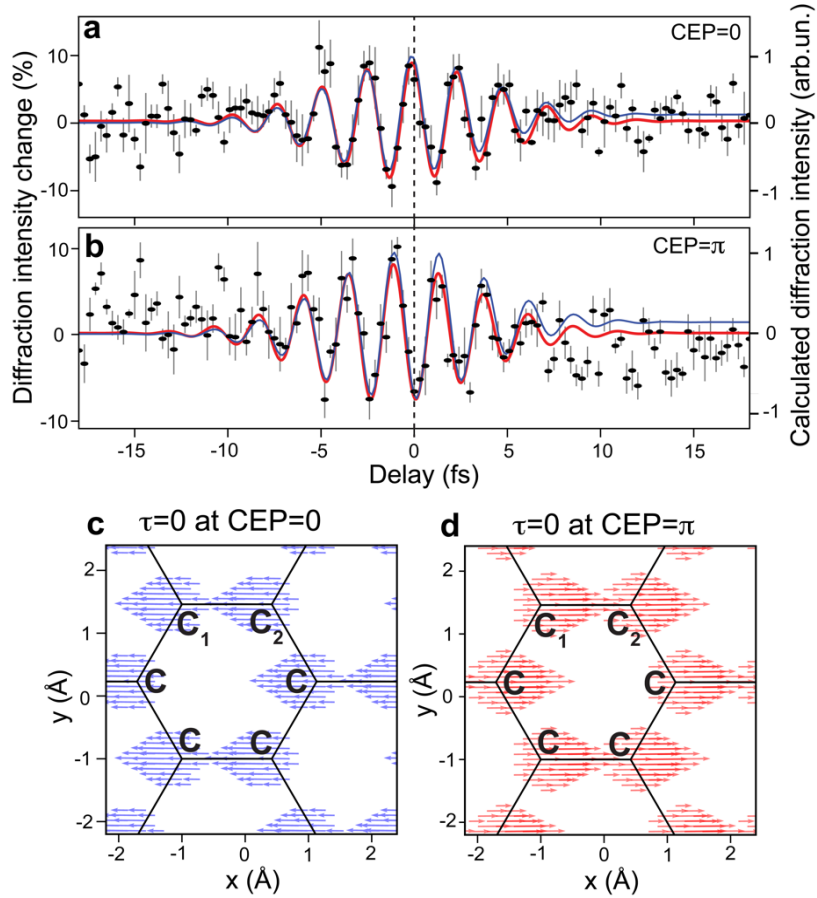

**Figure 3. Controlling the direction of graphene's electron motion in real space. (a & b) The measured (black dots and red solid lines, the error bars present are obtained from three scans) and calculated (blue solid lines) diffraction intensities of first-order Bragg peaks resulting from the interaction of the system with the driver fields with CEP~0 (a) and ~$\pi$ (b). (c &d) The generated electron currents were calculated at the same time delay ($\tau = 0$ fs) for the different CEP values.**

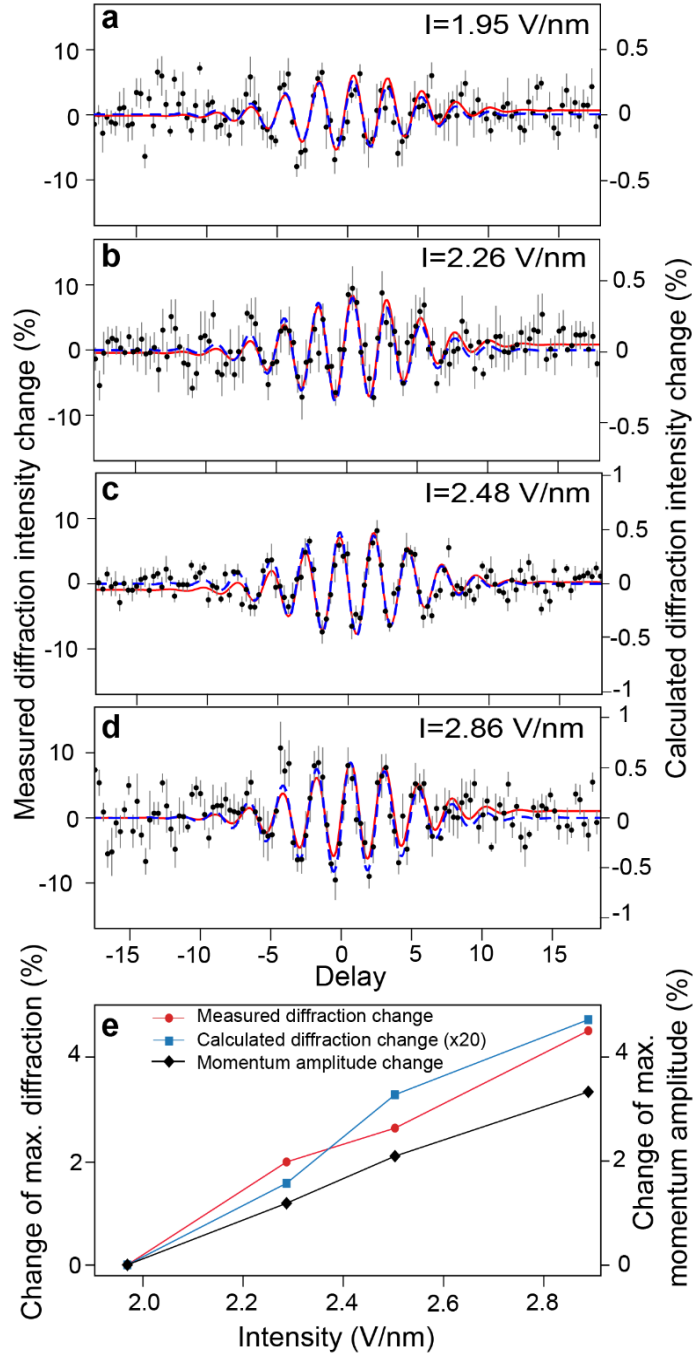

**Figure 4. Controlling the amplitude of graphene's electron motion. (a-d) The measured (black dots and red solid lines, the error bars are obtained from three scans) and calculated (blue solid lines) diffraction intensities in graphene were obtained as the result of interaction with the laser field of various intensities: 1.95, 2.26, 2.48, and 2.86 V/nm, respectively. (e) The relative changes as a function of the driver field strength in the maximum of the measured (red) and calculated (blue, multiplied by a factor of 20 for better visibility) diffraction intensities plotted in comparison with the relative changes in the averaged electron momentum.**

Second, the amplitude of the electron motion can be controlled by modifying the field intensity. To prove that experimentally, we conducted the diffraction measurements at different pump field strengths ranging from 1.95 to 2.86 V/nm (above 2.86 V/nm, we observed the sample damage). The field intensity was controlled by changing the power using a neutral density filter implemented in the beam pump beam. The measured diffraction intensities (black dots, fitted curves are shown by the red solid lines) are plotted in Fig. 4a-d together with the results of our simulations (blue solid lines). As one can see, we observed clear increase in the diffraction signals when raising the driver field strength. Figure 4e demonstrates relative linear growth in the maximum measured (red line) and calculated (blue line, multiplied by a factor of 20 for scaling) diffraction intensities as the function of the field strength. Both curves were obtained by normalizing the retrieved absolute values of the diffraction intensities by those observed at a field strength of 1.95 V/nm. The difference in slopes between the measured and calculated diffraction intensities results from the underestimation of the electron scattering by core electrons of graphene that are not considered in our simulations.

To quantify the observed changes in the scattering intensities and correlate them with the underlying electron motion, we calculated the averaged momentum of the excited electrons as a function of time:

$$\langle \mathbf{p}(t) \rangle = \frac{\int_{\text{unit cell}} \hbar \mathbf{k} \rho_{c,c}(\mathbf{k},t) d\mathbf{k}}{\int_{\text{unit cell}} \rho_{c,c}(\mathbf{k},t) d\mathbf{k}}, \qquad (5)$$

where $\rho_{c,c}(\mathbf{k},t)$ is the EMD in the conduction band of graphene, and the integration is performed within the unit cell. The relative changes in the maximum value of the electron momentum as a function of the field intensity are shown by the black solid line in Fig. 4e. Notably, both the measured scattering intensities and the averaged electron momentum show linear behaviour with a comparable increase of ~3 – 5% when measured at a field strength of 1.95 and 2.86 V/nm. Accordingly, we established the correspondence between the diffraction intensities, the generated electron current, and the averaged momentum of electrons, thus demonstrating the on-demand control of electron motion amplitude in graphene.

## 4. Conclusion

The attomicroscopy diffraction imaging permitted us to demonstrate experimentally the control of electron motion direction and amplitude in neutral multilayer graphene. We achieved this by changing the driver field parameters, namely the CEP and the intensity, and utilizing diffraction of attosecond gated electron pulses for the probe. Although already very successful, the anticipated further advancements of our attomicroscopy imaging setup will expand the power of this attosecond tool to demonstrate a vast range of real-life attosecond imaging applications. For instance, the potential development of energy-filtered imaging would directly connect the electron dynamics with the 3D structure of the sample under study, leading to the recording of movies of electron dynamics in real time and space. Consequently, full control of the ultrafast light fields and synthesized arbitrary waveforms by light field synthesis technology with attosecond resolution[19,34,35,39-51] would promise to control the material electronic properties for developing ultrafast optoelectronics[36,52,53]. Moreover, the on-demand electron motion control combined with the elementary mapping and electron-energy loss imaging options in the electron microscope would permit to control the chemical bonding formation and chemical reaction. Finally, the unique cryo-electron microscope imaging technology with the attosecond temporal resolution paves the way for the observation and manipulation of biological structures such as amino acids, DNA, and proteins.

## 5. Experimental Section/Methods

In the attomicroscopy setup, a few cycle NIR (800 nm) laser pulse is spitted to two beams. The small portion of the beam is used to generate a UV (400 nm) via second harmonic generation. This UV beam is sent to the electron microscope to generate an ultrafast electron pulse by photoemission from a photocathode. These electron pulses are accelerated to 200 KeV and sent to the gating medium and a sample where the optical gating takes place. The UV beam is kept at the lowest power to minimize the number of electrons per pulse and the related space-charge effect. Alternatively, the major portion of the NIR laser beam undergoes a strong nonlinear propagation through a hollow-core fibre (HCF) to generate a 5-fs laser pulse centred at 780 nm. This laser beam is divided by a beamsplitter into two beams. The first beam (pump beam) reflects off two mirrors mounted on a delay stage (piezo stage) with nanometre resolution to control the time delay. Note that neutral-density filters are introduced in the path of this beam to control the power. The second (gating) beam passes through a multiple-order quarter waveplate and then a zero-order quarter waveplate. This configuration allows us to generate a polarization-gated half-cycle laser pulse. This pulse is utilized as an optical gating pulse inside the microscope to generate the single attosecond electron pulse. Then, the gating and pump beams are combined and collinearly propagated and directed to the sample inside the microscope. The main ultrafast electron pulses are set in perfect temporal and spatial overlap with the gating laser pulse. Optical gating occurs on an aluminium TEM grid, and some of the electrons in the main electron pulse are gated in a time window equal to the linearly polarized half-cycle pulse of the gating pulse. The generated subfemtosecond electron pulses are used to image the electron motion dynamics triggered by the pump pulse in a single-crystal multilayer graphene sample by attosecond time-resolved electron diffraction measurements. In this experiment, the electron diffraction pattern of graphene is recorded as a function of the time delay (with a time step size of 300 as) between the gated electron and the pump pulses.

**Supporting Information**

Supporting Information is available from the Wiley Online Library or from the author.

**Acknowledgements**

This project is funded in part by the Gordon and Betty Moore Foundation Grant **GBMF 11476** to M. Hassan. Additionally, this material is based upon work partially supported by the Air Force Office of Scientific Research under award number **FA9550-22-1-0494**. We are also grateful to the W.M. Keck Foundation for supporting this project with a Science and Engineering award given to M.Th.H. Moreover, N.V.G. acknowledges support from the Branco Weiss Fellowship—Society in Science, administered by ETH Zürich.

# Attomicroscopy imaging and control of electron motion in graphene

*Mingrui Yuan [1,2][†], Husain Alqattan [1][†], Dandan Hui[1], Mohamed Sennary[1], Vladimir Pervak[3], Nikolay V. Golubev\* [1] and Mohammed Th. Hassan\* [1,2]*

[1] Department of Physics, University of Arizona, Tucson, AZ 85721, USA.

[2] James C. Wyant College of Optical Sciences, University of Arizona, Tucson, Arizona 85721, USA.

[3] Ludwig-Maximilians-Universität München, Am Coulombwall 1, 85748, Garching, Germany.

[†] These authors contributed equally to this work

*Corresponding author e-mail: mohammedhassan@arizona.edu and ngolubev@arizona.edu

## 1. Fitting methodology

To retrieve a continuous diffraction signal from the discrete data sets obtained in our experiments, we fitted our data by a Gaussian-modulated cosine function, mathematically described as:

$$f(t) = A\exp\left(-t^2/2\sigma^2\right)\cos(\omega t + \varphi) + C. \tag{SI6}$$

In this expression:

- *A* represents the amplitude,
- *σ* defines the width of the Gaussian,
- *ω* is the angular frequency,
- *φ* is the carrier-envelope phase,
- *C* is the offset that accounts for any baseline shift in the measured data.

In the fitting of the diffraction intensities, the angular frequency $\omega$ was assumed to be identical to that of the driver field (1.65 eV photon energy). Similarly, we assumed the duration of the scattering response to be identical to the driver field duration (FWHM = $2\sqrt{2\ln 2}\sigma$ = 11.209 fs). Furthermore, all experimental data were shifted in time such that the maximum measured diffraction intensity was centred at zero time delay for convenience. The three parameters (*A*, *φ*, and *C*) have been optimized in the fitting process. A maximum number of 50,000 fitting steps were allowed to ensure the convergence. The obtained parameters for all the experimental data presented in the paper are given in Table 1.

|  | El. field strength (V/nm) | $A$ (arb. units) | $\varphi$ ($\pi$) | $C$ (arb. units) |
| --- | --- | --- | --- | --- |
| Figure 2a | 2.483 | 0.080 | 0.539 | 1.002 |
| Figure 3a | 2.860 | 0.084 | -0.184 | 1.015 |
| Figure 3b | 2.483 | 0.080 | 0.885 | 1.033 |
| Figure 4a | 1.950 | 0.051 | -0.357 | 1.001 |
| Figure 4b | 2.260 | 0.080 | -0.335 | 0.995 |
| Figure 4c | 2.483 | 0.080 | 0.539 | 1.002 |
| Figure 4d | 2.860 | 0.081 | -0.184 | 1.015 |

**Table 1. Parameters obtained by fitting the Gaussian-modulated cosine function, Eq. (SI6), to the experimental data presented in the main text.**